\documentclass[cernpreprint, atlasdraft=false, USenglish, cmslogo=true, coverpage=false, texmf, orcidlogo]{atlasdoc}
\usepackage{atlaspackage}
\usepackage{atlasbiblatex}

\graphicspath{{logos/}{figures/}{figures/results/Paper/}}

\usepackage{ANA-HIGP-2025-18-PAPER-defs}

\usepackage{hepparticles}
\usepackage{heppennames2}
\usepackage{ptdr-definitions}
\usepackage{multirow}

\newcommand{\HH}{\ensuremath{{\PH\PH}}\xspace}

\renewcommand{\bbbar}{\ensuremath{\PQb\PQb}\xspace}

\newcommand{\bbgg}{\ensuremath{\bbbar\gamma\gamma}\xspace}

\newcommand{\bbbb}{\ensuremath{\bbbar\bbbar}\xspace}
\newcommand{\bbtt}{\ensuremath{\bbbar\PGt\PGt}\xspace}

\newcommand{\bbWW}{\ensuremath{\bbbar\PW\PW}\xspace}
\newcommand{\WWgg}{\ensuremath{\PW\PW\gamma\gamma}\xspace}
\newcommand{\ggtt}{\ensuremath{\PGt\PGt\gamma\gamma}\xspace}
\newcommand{\bbZZfourl}{\ensuremath{\bbbar\PZ\PZ(4\ell)}\xspace}

\newcommand{\HHbbgg}{\ensuremath{\bbbar\gamma\gamma}\xspace}
\newcommand{\HHbbtt}{\ensuremath{\bbbar\PGt\PGt}\xspace}
\newcommand{\HHbbWW}{\ensuremath{\bbbar\PW\PW}\xspace}
\newcommand{\HHbbZZ}{\ensuremath{\bbbar\PZ\PZ}\xspace}
\newcommand{\HHbbll}{\ensuremath{\bbbar\ell\ell+E_{\mathrm{T}}^{\text{miss}}}\xspace}

\newcommand{\HHbbbb}{\ensuremath{\bbbar\bbbar}\xspace}
\newcommand{\HHLeptons}{\ensuremath{\text{multilepton}}\xspace}

\newcommand{\Hgg}{\ensuremath{\PH\to\gamma\gamma}\xspace}
\newcommand{\Hbb}{\ensuremath{\PH\to \bbbar}\xspace}

\newcommand{\Htt}{\ensuremath{\PH\to \tau \tau}\xspace}

\newcommand{\tauhad}{\ensuremath{\PGt_{{\text h}}}\xspace}

\newcommand{\mHH}{\ensuremath{m_{\PH\PH}}\xspace}

\newcommand{\PTH}{\ensuremath{p_{\mathrm{T}}^{\PH}}\xspace}

\providecommand*\klambda{\ensuremath{\kappa_\lambda}\xspace} %

\providecommand*\ktop{\ensuremath{\kappa_{\PQt}}\xspace} %
\renewcommand{\kt}{\ktop}

\newcommand{\kVV}{\ensuremath{\kappa_{2\PV}}\xspace}

\newcommand{\kvv}{\ensuremath{\kVV}\xspace}

\providecommand{\HH}{\ensuremath{\PH\PH}}
\providecommand{\klambda}{\ensuremath{\kappa_\lambda}}
\providecommand{\kvv}{\ensuremath{\kappa_{2\text{V}}}}
\providecommand*{\fb}{\mbox{\ensuremath{\,\text{fb}}}\xspace} %
\providecommand*{\pb}{\mbox{\ensuremath{\,\    text{pb}}}\xspace} %
\providecommand{\muHHnull}{\ensuremath{\mu_\HH^\text{Asimov}}\xspace}

\AtlasTitle{Combination of ATLAS and CMS searches for Higgs boson pair production at $\sqrt{s} = 13$~TeV}

\AtlasAbstract{%

This Letter presents a combination of searches for Higgs boson pair (HH) production performed by the ATLAS and CMS Collaborations using proton--proton collision data sets recorded at  $\sqrt{s} = 13$~TeV during the Large Hadron Collider Run~2, corresponding to integrated luminosities ranging between 126 and 140 $\mathrm{fb^{-1}}$.
The upper limit at the 95\% confidence level on the total HH signal strength, defined as the ratio of the measured cross section to the SM prediction, corresponds to 2.5, with an expected value of 1.7 (2.8) assuming the absence (presence) of the standard model (SM) HH signal.
The strength of the HH signal is measured to be
$0.8^{+0.9}_{-0.7}$
relative to the SM prediction.
The observed significance is found to be 1.1 standard deviations whereas 1.3 are expected for the SM HH signal.
Constraints are set on the Higgs boson trilinear self-coupling and on the couplings of two Higgs bosons to two vector bosons, both normalized to the SM predictions and denoted as $\kappa_\lambda$ and $\kappa_{2\mathrm{V}}$, respectively.
The observed individual constraints at the 95\% confidence level are
$-0.71 < \kappa_\lambda < 6.1$ and
$0.73 < \kvv < 1.3$,
while the expected constraints assuming the presence of the SM HH signal are
$-1.3 < \kappa_\lambda < 6.7$ and
$0.66 < \kappa_{2\mathrm{V}} < 1.4$.

}

\AtlasRefCode{HIGP-2025-18}

\PreprintIdNumber{CERN-EP-2026-011}

\AtlasJournal{Phys.\ Rev.\ Lett.}
\AtlasCoverEgroupAnalysisTeam{atlas-HIGP-2025-18-analysis-team@cern.ch}

\hypersetup{pdftitle={ATLAS and CMS document},pdfauthor={The ATLAS anc CMS Collaborations}}

\begin{document}

\maketitle


The standard model of particle physics (SM)~\cite{StandardModel67_1,StandardModel67_2,StandardModel67_3} includes a doublet of complex scalar fields, $\Phi$, that is subject to a potential which is symmetric under the $SU(2) \times U(1)$ electroweak gauge group.
This potential has the form
$V(\Phi^\dagger\Phi) = -\mu^2 \Phi^\dagger\Phi + \lambda (\Phi^\dagger\Phi)^2$, with parameters $\mu^2>0$ and $\lambda > 0$ .
In the mechanism postulated by Brout, Englert, Higgs, and others~\cite{Englert:1964et,Higgs:1964ia,Higgs:1964pj,Guralnik:1964eu}, the electroweak symmetry is spontaneously broken when $\Phi$ acquires a non-zero vacuum expectation value $v=|\mu|/\sqrt{\lambda} \approx 246\GeV$, giving rise to the longitudinal polarizations of the $\PW$ and $\PZ$ vector bosons and resulting in a physical scalar field $h$ whose associated quantum excitations are Higgs bosons (\PH).
In the SM, the Higgs boson mass $m_\PH$ depends on the curvature of the potential, $m_\PH^2 = 2\lambda v^2$.
The SM also predicts the existence of a Higgs boson trilinear self-coupling ($\PH\PH\PH$) in the form $\lambda v h^3$ and of a coupling between two Higgs bosons and two vector bosons ($\PV\PV\PH\PH$, with $\PV = \PW,\PZ$), whose strength depends on $(m_\PV/v)^2$, where $m_\PV$ is the $\PV$ boson mass, and is directly proportional to the strength of the coupling between one Higgs and two vector bosons ($\PV\PV\PH$).
The spontaneous breaking of the electroweak symmetry and the resulting properties of the scalar sector are thus determined by the form of the potential via the value of $\lambda$.

The existence of the Higgs boson was established by the ATLAS and CMS Collaborations in 2012~\cite{ATLAS:2012yve,CMS:2012qbp,CMS:2013btf}.
Since then, all measurements performed to characterize its properties have yielded observations compatible with the SM predictions~\cite{HIGG-2021-23,CMS-HIG-22-001}.
These measurements inform us about the behavior of the $h$ field near the minimum of the potential.
However, the global structure of the potential, which in the SM is controlled by the value of $\lambda$, is largely unknown.
Its study represents one of the main goals of the CERN Large Hadron Collider (LHC)~\cite{LHCReference} physics program.

The Higgs boson trilinear self-coupling, whose strength depends on $\lambda$, can be directly measured through the production of Higgs boson pairs (\HH). In the SM, the leading \HH production mode in proton--proton ($\Pp\Pp$) collisions at the LHC is gluon fusion (ggF), with a cross section of
$\sigma^\text{ggF}_\HH = 30.8^{+2.0}_{-7.1}\fb$~\cite{Dawson:1998py,PhysRevLett.117.012001,Baglio:2018lrj,deFlorian:2013jea,Shao:2013bz,deFlorian:2015moa,Grazzini:2018bsd,Baglio:2020wgt}, followed by vector boson fusion (VBF) with a cross section $\sigma^\text{VBF}_\HH = 1.687 \pm 0.046 \fb$~\cite{Baglio:2012np,Frederix:2014hta,Ling:2014sne,Dreyer:2018rfu,Dreyer:2018qbw,Dreyer:2020xaj}. %
Both predictions are computed for a center-of-mass energy of 13\TeV and for $m_{\PH} = 125\GeV$, using the PDF4LHC21 parton distribution function (PDF) set~\cite{PDF4LHCWorkingGroup:2022cjn}.
These processes exhibit a quadratic dependence of their total cross section at the leading electroweak order on the value of the Higgs boson trilinear self-coupling relative to the SM prediction, which is referred to as $\klambda$.
Modifications of $\klambda$ also lead to non-trivial kinematic changes of the \HH processes because of the interference between the contributions involving the Higgs boson trilinear self-coupling and the ones only involving other Higgs boson interactions~\cite{Carvalho:2015ttv,Capozi:2019xsi}.
The total ggF cross section exhibits a minimum for $\klambda \approx 2.3$,
where the interference effects are strongest~\cite{Amoroso:2020lgh,Heinrich:2022idm,Bagnaschi:2023rbx}.
Large values of \klambda result in an enhancement of the cross section in the region with low \HH invariant mass ($\mHH$), close to the kinematic production threshold.
The VBF production cross section also features a quadratic dependence on the strength of the $\PV\PV\HH$ interaction, whose value normalized to the SM prediction is referred to as $\kvv$.
A single parameter is considered in this work to scale simultaneously the $\PW\PW\PH\PH$ and $\PZ\PZ\PH\PH$ couplings, since these are not distinguished experimentally in VBF \HH analyses.
Scenarios with $\kvv\neq1$ are characterized by a strong increase in the signal cross section at large values of $m_\HH$ and the resulting production of Higgs bosons with large transverse momenta (\PTH)~\cite{Bishara:2016kjn}, a phase space region offering enhanced experimental sensitivity.

Both the ATLAS and CMS Collaborations have performed searches for \HH production using the $\Pp\Pp$ Run-2 data sets collected at $\sqrt{s}=13\TeV$ between 2015 and 2018, corresponding to integrated luminosities ranging between 126 and 140\fbinv, depending on the experiment and the analysis considered.
Searches have been performed by both experiments in a variety of final states~\cite{Tumasyan:2801882,HDBS-2019-29,arXiv:2205.06667, HDBS-2022-02,arXiv:2206.09401, HDBS-2019-27,Sirunyan:2745738,HDBS-2021-10,CMS-HIG-21-005,HDBS-2019-02,CMS:2025tqi,CMS:2022omp,CMS:2024fkb} and combined independently by the two collaborations.
In terms of the total signal strength $\mu_\HH$, defined as the ratio of the measured and predicted values of $\sigma^\text{ggF}_\HH + \sigma^\text{VBF}_\HH$, the observed (expected) upper limits assuming the absence of \HH production are 2.9 (2.4) for ATLAS~\cite{ATLAS:2024ish} and  3.5 (2.5) for CMS~\cite{CMS_combination} at the 95\% confidence level (CL).

This Letter presents the combination of ATLAS and CMS searches for \HH production using the LHC Run-2 data sets. The input analyses correspond to the
\bbbb~\cite{Tumasyan:2801882,HDBS-2019-29,arXiv:2205.06667, HDBS-2022-02}, \bbWW~\cite{CMS-HIG-21-005,HDBS-2019-02}, \bbtt~\cite{arXiv:2206.09401, HDBS-2019-27}, \bbgg~\cite{Sirunyan:2745738,HDBS-2021-10}, and \HHLeptons~\cite{CMS:2022kdx, HDBS-2019-04} searches, where the latter refers collectively to topologies with multiple leptons.
Charge conjugation is implied in the notation used throughout this Letter.
These searches, as included in the CMS and ATLAS only combinations, have the leading sensitivity to SM \HH production, \klambda, and \kvv in both experiments. The ATLAS \HHLeptons search targets additional final states such as $\bbZZfourl$, with $\ell = \Pe, \PGm$, and signatures with both multiple leptons and photons. CMS covers this additional phase space with dedicated searches in $\ggtt$~\cite{CMS:2025tqi}, $\bbZZfourl$~\cite{CMS:2022omp} and $\WWgg$~\cite{CMS_combination}. These CMS analyses as well as the \HHbbWW (hadronic)~\cite{CMS_combination} and VHH ($\HH\rightarrow\HHbbbb$)~\cite{CMS:2024fkb} searches originally included in the CMS combination are not considered for this work. This choice reduces the technical complexity of the combination with an expected impact of about 1\% on the combined sensitivity to the SM \HH signal strength.
The included analyses and their relative sensitivities are summarized in Table~\ref{tab:inputnalysis}.

The ATLAS~\cite{ATLAS:2008xda} and CMS~\cite{CMS:2008xjf,CMS:2023gfb} experiments are the two multipurpose particle detectors at the LHC.
Both detectors have a nearly hermetic design with a forward--backward symmetric cylindrical geometry, and are designed to reconstruct and identify electrons, muons, photons, and hadronic objects that are used as inputs to the analyses.
The ATLAS detector consists of an inner tracking detector based on silicon pixel, silicon microstrip, and transition radiation detectors.
It is surrounded by a thin superconducting solenoid providing a 2\,T axial magnetic field, lead/liquid-argon sampling electromagnetic calorimeters, steel/scintillator-tile and liquid-argon hadronic calorimeters with lead, tungsten, or copper absorbers, and a muon spectrometer based on superconducting air-core toroidal magnets.
The CMS apparatus features a 3.8\,T superconducting solenoid of 6\,m internal diameter.
An all-silicon inner tracker, a lead tungstate crystal electromagnetic calorimeter, and a brass/scintillator hadron calorimeter are contained within the solenoid volume, and gas-ionization muon detectors are embedded in the flux-return yoke outside the solenoid.

The \HHbbbb final state has a branching fraction of about 34\%~\cite{LHCHiggsCrossSectionWorkingGroup:2016ypw}, the largest among the \HH decay modes, but faces a challenging multijet background.
The selection of signal events relies heavily on the use of modern heavy-flavor jet taggers and dedicated $\PQb$-jet trigger algorithms.
Both experiments target resolved-jet~\cite{Tumasyan:2801882,HDBS-2019-29} and merged-jet~\cite{arXiv:2205.06667, HDBS-2022-02} signatures, where the $\PH\to\bbbar$ decays are respectively reconstructed as two separate jets with a radius parameter $R = 0.4$, or as a single large-radius jet with $R = 0.8$ for CMS and $R = 1$ for ATLAS.
In both cases, jets are reconstructed with the anti-\kt algorithm~\cite{Cacciari:2008gp,Cacciari:2011ma}.
The merged-jet topology has leading sensitivity to VBF \HH production in anomalous $\kvv$ scenarios where the signal is characterized by large $m_\HH$ and large \PTH.
The ATLAS merged-jet analysis only searches for VBF \HH production with dedicated selections on additional jets in the event, while the CMS merged-jet analysis covers both ggF and VBF \HH production modes.
All four analyses estimate background contributions, mostly consisting of multijet production, using data in background enriched control regions, and simulation for the remaining processes in the merged-jet CMS analysis.
The final results are extracted using a template fit either to a boosted decision tree (BDT) classifier output or the reconstructed \mHH and merged-jet mass distributions.

The \HHbbtt final state combines a sizable total branching fraction of about 7.3\%~\cite{LHCHiggsCrossSectionWorkingGroup:2016ypw} with reasonable selection purity from the identification of the $\PGt\PGt$ decay products.
Both ATLAS and CMS analyses~\cite{arXiv:2206.09401, HDBS-2019-27} target the decay channels of the $\PGt\PGt$ system where at least one $\PGt$-lepton decays to hadrons and a neutrino ($\tauhad$).
Events are classified in a variety of signal regions based on the final state and the corresponding single-lepton, di-$\tauhad$, and lepton + $\tauhad$ triggers (ATLAS) or on the presence of one or two small-radius $\PQb$-tagged jets or a single large-radius jet (CMS).
Categories enriched in VBF \HH events are also defined.
The results are extracted using BDT or deep neural network (DNN) classifier scores for the ATLAS and CMS searches, respectively.
Templates for the distributions of the background are mostly derived from simulation, and control regions in data are used to model the contribution from misidentified $\tauhad$ objects.

The \HHbbgg final state compensates an extremely low branching fraction of 0.26\%~\cite{LHCHiggsCrossSectionWorkingGroup:2016ypw} with a clean experimental signature from the $\PH\to\gamma\gamma$ decay.
The signal extraction strategies in the ATLAS and CMS analyses~\cite{Sirunyan:2745738,HDBS-2021-10} are similar and feature a parametric fit to the data in the diphoton invariant mass (ATLAS and CMS) and the dijet mass observable (CMS).
These fits are performed in a number of subcategories of different signal purity. Categories are built based on the modified four-body invariant mass $m_{\PQb\PQb\gamma\gamma}^{*}=m_{\PQb\PQb\gamma\gamma}- (m_{\PQb\PQb} - 125 \,\GeV) - (m_{\gamma\gamma} -125 \,\GeV)$ and BDT outputs separating the signal from the continuum diphoton background and, in the case of ATLAS, also from single \PH processes.
For ATLAS a BDT is separately trained in the high- and low-$m_{\PQb\PQb\gamma\gamma}^{*}$ regions.
In the CMS analysis, additional categories targeting VBF \HH production are used, and the resonant single \PH background is reduced by applying a selection requirement based on a dedicated multivariate analysis classifier (MVA) in the ggF categories.

The \HHbbWW decay channel offers the second-highest branching fraction of all \HH decay modes at about 25\%~\cite{LHCHiggsCrossSectionWorkingGroup:2016ypw}. The CMS \HHbbWW~\cite{CMS-HIG-21-005} and
ATLAS \HHbbll~\cite{HDBS-2019-02}
analyses target this final state in semi-leptonic (CMS) and dileptonic (ATLAS and CMS) \HHbbWW final states, with additional contributions from \HHbbtt and \HHbbZZ decays.
The final state leptons ($\ell$) considered are electrons and muons, and the symbol $E_{\mathrm{T}}^{\text{miss}}$ denotes the magnitude of the missing transverse momentum.
In the CMS case, events in the semi-leptonic and dileptonic categories are further split based on the number of $\PQb$-tagged jets, their reconstruction as a merged large-radius jet or as resolved small-radius jets and the output of a multiclass DNN separating ggF \HH, VBF \HH, $\PQt+\PH$, and other backgrounds. For ATLAS, events are divided into signal and control regions based on the presence of a ggF- or VBF-like topology, $m_{\ell\ell}$, $m_{\PQb\PQb}$, and the lepton flavor. For the ggF-like case a multivariate DNN is trained to classify events into signal, $\PQt\PAQt+\PW\PQt$, and other backgrounds, while in the VBF-like case a BDT is used to separate the signal from backgrounds. In both the ATLAS and CMS analyses, the signal is extracted from the MVA discriminator outputs of the various categories.

Analyses targeting final states with multiple leptons (electrons, muons, $\tauhad$) in combination with other objects benefit from high trigger efficiencies and low reconstruction thresholds, resulting in large acceptances for signal hypotheses with soft \mHH spectra and low \PTH.
The CMS \HHLeptons analysis~\cite{CMS:2022kdx} targets the $\PW\PW\PW\PW$, $\PW\PW\PGt\PGt$, and $\PGt\PGt\PGt\PGt$ decay channels, while the ATLAS analysis~\cite{HDBS-2019-04} additionally targets the decays to $\PQb\PQb\PZ\PZ$, and to a $\gamma\gamma$ pair plus a $\PW\PW,\,\PZ\PZ,$ or a $\tau\tau$ pair. The total branching fraction of all these decay modes is about 9\%~\cite{LHCHiggsCrossSectionWorkingGroup:2016ypw}, with the various final states included in the ATLAS and CMS analyses corresponding to a branching fraction of about 5\% due to lepton requirements.
The signal is extracted in categories based on how many light lepton, $\tauhad$, and photon (ATLAS) objects are present. In each category, a fit is performed to the output of dedicated BDTs or to the diphoton invariant mass for $\gamma\gamma$ final states.
Signal and background templates are obtained from a combination of estimations based on simulation and control regions in data.

The overlap between the selection criteria of the input analyses within each experiment was studied as part of the individual combinations.
It was found to be negligible and treated as described in Refs.~\cite{ATLAS:2024ish,CMS_combination}.

Signals in the input analyses are normalized to the predictions based on the PDF4LHC15 PDF set~\cite{Butterworth:2015oua}.
Compared to the most recent cross section calculations, based on the PDF4LHC21 PDF set and reported earlier, the predictions used for this work are less than 1\% larger for the ggF production mode, independent of the \klambda hypothesis.
For the VBF production mode, the predictions used for this work are about 2\% larger compared to the most recent ones, which use the updated PDF set and include next-to-leading order electroweak corrections~\cite{Dreyer:2020xaj}.
The corresponding total SM cross section for \HH production is $32.8\fb$.
The ggF \HH signal prediction~\cite{Heinrich:2017kxx,Heinrich:2019bkc}, for each of the analyses used in this combination, is rescaled by a function of \klambda to adjust its normalization following the correction of the two-loop amplitude in the \textsc{POWHEG}~\cite{Frixione:2007vw,Alioli:2010xd} simulation described in Refs.~\cite{Heinricherratum,Bagnaschi:2023rbx}.
The effect of this modification on the distributions of simulated \HH events is not taken into account in the ATLAS and CMS input analyses, but its impact was estimated~\cite{ATLAS:2024ish} to be less than 4\% on the constraints for \klambda, and the SM result is not affected.

The dependence of the single \PH total production cross sections and branching fractions on \klambda from loop-level effects~\cite{Degrassi:2016wml,Maltoni:2017ims} is included in the ATLAS \HHbbtt and in all CMS input analyses through a parametrization of the total cross sections of single \PH background processes as a function of \klambda.
The same parametrization was added for this Letter to the background modeling of the ATLAS \HHbbgg and \HHbbbb analyses where it was not originally implemented.
These effects are not considered for the ATLAS analyses in the \HHLeptons and \HHbbll final states, because their statistical models do not contain a parametrization of individual single \PH processes, preventing the addition of such \klambda dependence.
Considering that their implementation in the ATLAS \HHbbgg and \HHbbbb input analyses has an impact smaller than 1\% on the combined \klambda sensitivity, the effects from the missing corrections in the ATLAS \HHLeptons and \HHbbll inputs on the overall inference is estimated to be negligible.

Results are obtained using a profile likelihood ratio test statistic with systematic uncertainties modeled as nuisance parameters and making use of the asymptotic approximation~\cite{Cowan:2010js}.
The combined likelihood is the product of the individual likelihoods and the signal predictions are parametrized by the \HH signal strength parameter, $\mu_\HH$, and by the \klambda and \kvv parameters, common to both experiments.
In fitting to the data either the signal strength or the coupling modifiers are allowed to float, but not both.
Upper limits on $\mu_\HH$ are computed using the modified frequentist $\text{CL}_s$ criterion~\cite{Junk:1999kv,Read:2002hq} with the profile likelihood ratio modified for upper limits~\cite{Cowan:2010js,CMS:2024onh} as the test statistic.
Asimov datasets~\cite{Cowan:2010js} are generated by setting all nuisance parameters to their best-fit value and $\mu_\HH$, \klambda, and \kvv according to the assumed hypothesis, and are used to compute expected results.
Results are derived and cross-checked by the two collaborations using different software.
The ATLAS Collaboration uses \textsc{HistFactory}~\cite{Cranmer:1456844} and the CMS Collaboration uses \textsc{Combine}~\cite{CMS:2024onh}, both of which are based on the \textsc{RooFit}~\cite{verkerke2003roofittoolkitdatamodeling} and \textsc{RooStats}~\cite{moneta2011roostatsproject} packages.

Theoretical uncertainties in the \HH signal production cross sections are related to the missing orders in the QCD calculations, estimated by varying the factorization and renormalization scales, to the choice of the top quark mass scheme, and to the limited knowledge of the PDFs and $\alpha_S$.
These uncertainties are considered as fully correlated across all the input analyses and the two experiments.
Since they affect theoretical predictions, they are not considered when computing upper limits on the \HH production cross section, while they are accounted for when deriving the other results.
Similarly, the corresponding sources of uncertainties on the single \PH processes are correlated between the two experiments.
Such uncertainties are related to the cross section of \PH production via ggF, via VBF, and in association with top quarks, to the normalization of ggF and VBF \PH production in association with heavy flavor jets, and to the branching fractions for \Hbb, \Htt, and \Hgg decays.
Uncertainties for the other decay modes are considered but not correlated across experiments since they follow a different grouping because of the final states considered in the \HHbbWW and \HHLeptons analyses.
All other uncertainties considered in the input analyses are not correlated between the two experiments, while correlations across different analyses from the same experiment follow their original treatment~\cite{ATLAS:2024ish,CMS_combination}.
Scenarios with a correlated uncertainty in the integrated luminosity and with a different correlation scheme between groups of single \PH processes have been studied and were found to have a negligible effect on the results.
The magnitude of the uncertainty in the normalization of single \PH production in association with heavy flavor jets~\cite{Manzoni:2023qaf} is 100\% for ATLAS analyses and 50\% for CMS analyses.
The choice to retain two different values of this uncertainty follows from the small impact in the present results and for consistency with previously published results~\cite{ATLAS:2024ish,CMS_combination} that are used as input to this work.
Considering a common correlated uncertainty of 100\% for both experiments increases the observed 95\% CL upper limit on the \HH signal strength by 1\% without  significant change to the expected limit, and has similar effects at the percent level or below on the best-fit value of $\mu_\HH$ and the 95\% CL constraints on \klambda.
These uncertainties are treated as fully correlated between the two experiments, while the change in the $\mu_\HH$ and \klambda sensitivities if they are considered as uncorrelated is about 1\%.

The individual per-experiment best-fit values on the total signal strength $\mu_\HH$, obtained from individual fits to each set of results, are
$\hat{\mu}_\HH^\text{ATLAS}=0.5^{+1.2}_{-1.1}$ and
$\hat{\mu}_\HH^\text{CMS}=1.0^{+1.3}_{-1.0}$.
The combined best-fit signal strength for \HH production is found to be
$\hat{\mu}_\HH =
0.8^{+0.9}_{-0.7} \allowbreak =
0.8^{+0.7}_{-0.6} (\text{stat.}) \allowbreak{}^{+0.4}_{-0.2} (\text{theory}) \allowbreak{}^{+0.3}_{-0.3} (\text{exp.}) \allowbreak
$,
where the breakdown of the uncertainty is into the statistical, theoretical, and experimental components, computed as detailed in Ref.~\cite{ATLAS:2016neq}.
The expected best-fit value is
$\hat{\mu}_\HH^\text{exp} =
1.0^{+0.9}_{-0.8}  \allowbreak =
1.0^{+0.7}_{-0.6} (\text{stat.})  \allowbreak{}^{+0.5}_{-0.3} (\text{theory})  \allowbreak{}^{+0.3}_{-0.3} (\text{exp.})
$.
The leading theoretical uncertainties are those on the ggF \HH signal normalization from the choice of the renormalization scheme and scale of the top quark mass, and from the production of single \PH in association with heavy flavor jets, the latter impacting mostly the \HHbbgg and \HHbbtt analyses.
The leading experimental uncertainties are related to the background modeling in the CMS \bbbb merged-jet and ATLAS \bbtt analyses.
The combined observed (expected) significance for the \HH signal is 1.1 (1.3) standard deviations.

\begin{figure}[h]
\centering
\includegraphics[width=0.6\linewidth]{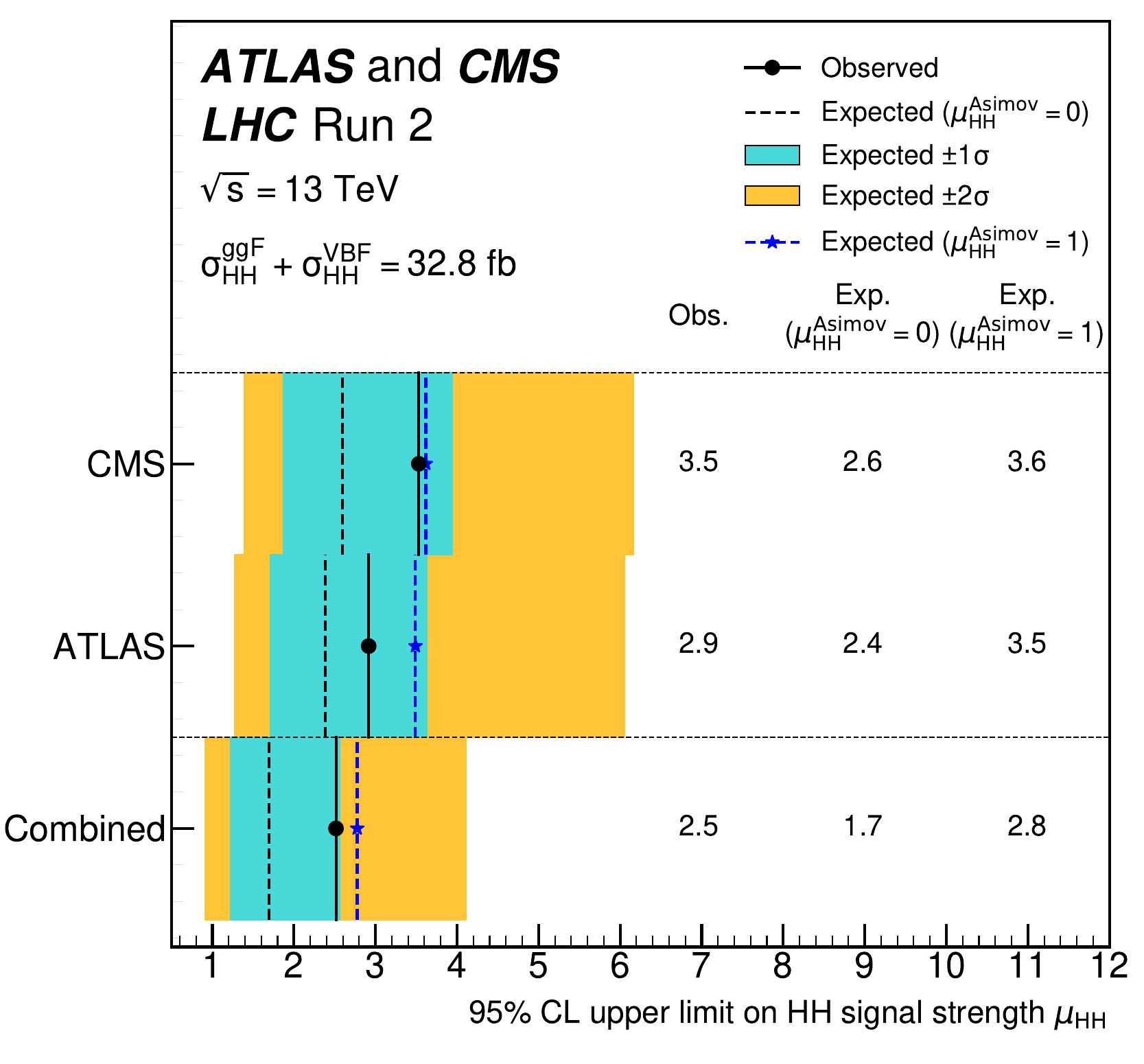}
\caption{Expected and observed 95\% CL upper limits on the total \HH signal strength, defined as the ratio of the measured cross section to the sum of the ggF and VBF \HH SM cross sections, for ATLAS, CMS, and the combination from both experiments. The median expected limits on $\mu_\HH$ are obtained under the hypotheses of no \HH signal ($\muHHnull=0$) or assuming the presence of the SM \HH signal ($\muHHnull=1$). The $\pm 1\sigma$ and  $\pm 2\sigma$ bands are computed under the $\muHHnull=0$ hypothesis.
}
\label{fig:smlimits}
\end{figure}

Upper limits are placed on the total \HH production cross section
assuming the absence of \HH production ($\muHHnull = 0$).
The observed combined limit at 95\% CL corresponds to $73\fb$ with an expected limit of 50~\fb.
Taking into account uncertainties in the \HH signal cross section, this translates into an observed 95\% CL upper limit on the \HH signal strength of 2.5, with an expected limit of 1.7, as shown in Figure~\ref{fig:smlimits}.
With limits approaching the SM range, it can be insightful to also compare with the expected upper limit computed assuming SM \HH production ($\muHHnull = 1$), which is 2.8 times the SM prediction, slightly closer to the observed limit.
An upper limit on the individual signal strength of the VBF \HH production mode, $\mu_\HH^\text{VBF}$, is obtained by fixing the ggF \HH signal to the SM prediction.
The observed combined limit at 95\% CL on $\mu_\HH^\text{VBF}$ is 35 with an expected value of 41 assuming the absence of the VBF \HH signal.

\begin{figure}[h]
\centering
\subfloat[]{
\includegraphics[width=0.49\linewidth]{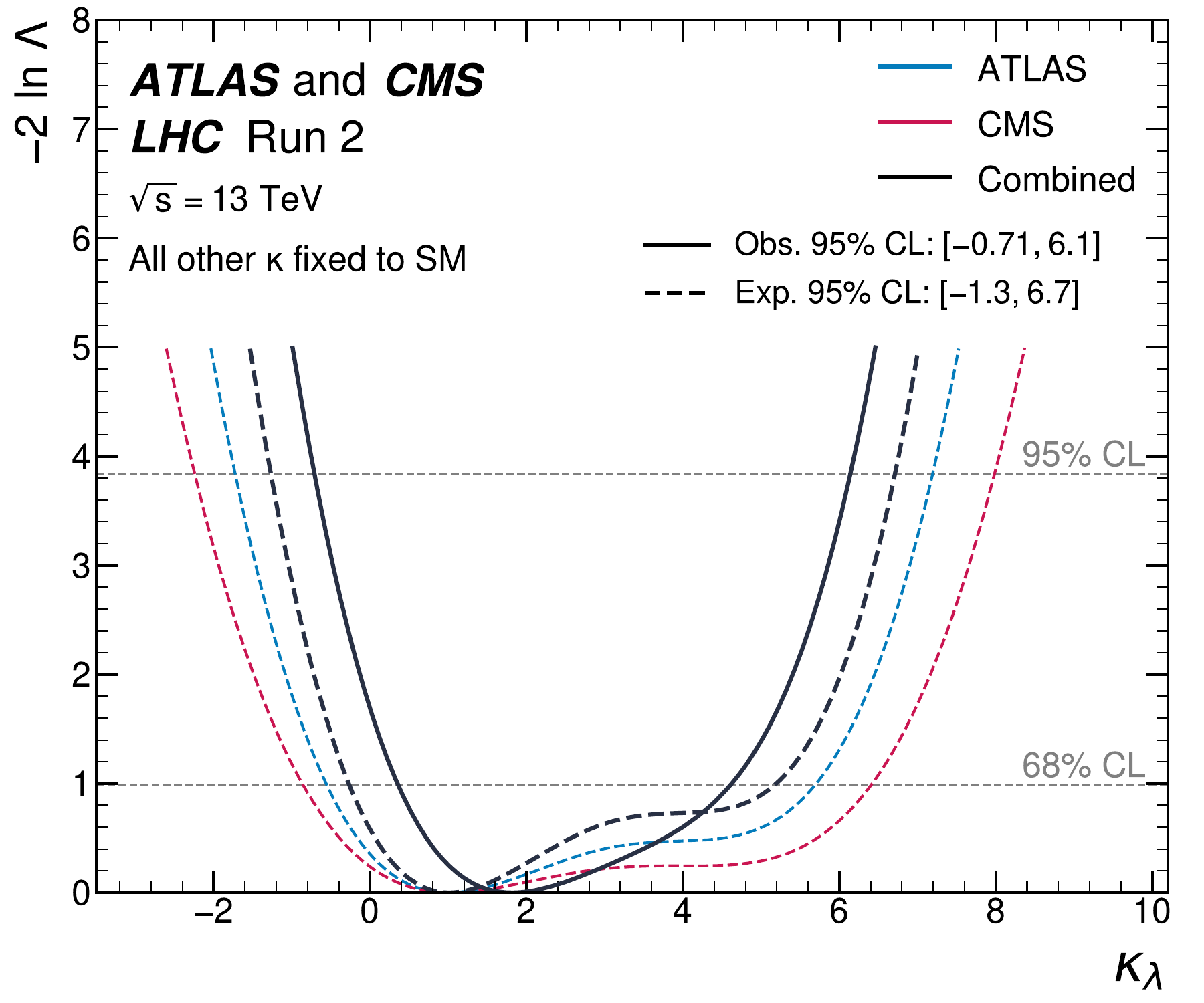}
\label{fig:likelihoodscan:klambda}
}
\subfloat[]{
\includegraphics[width=0.49\linewidth]{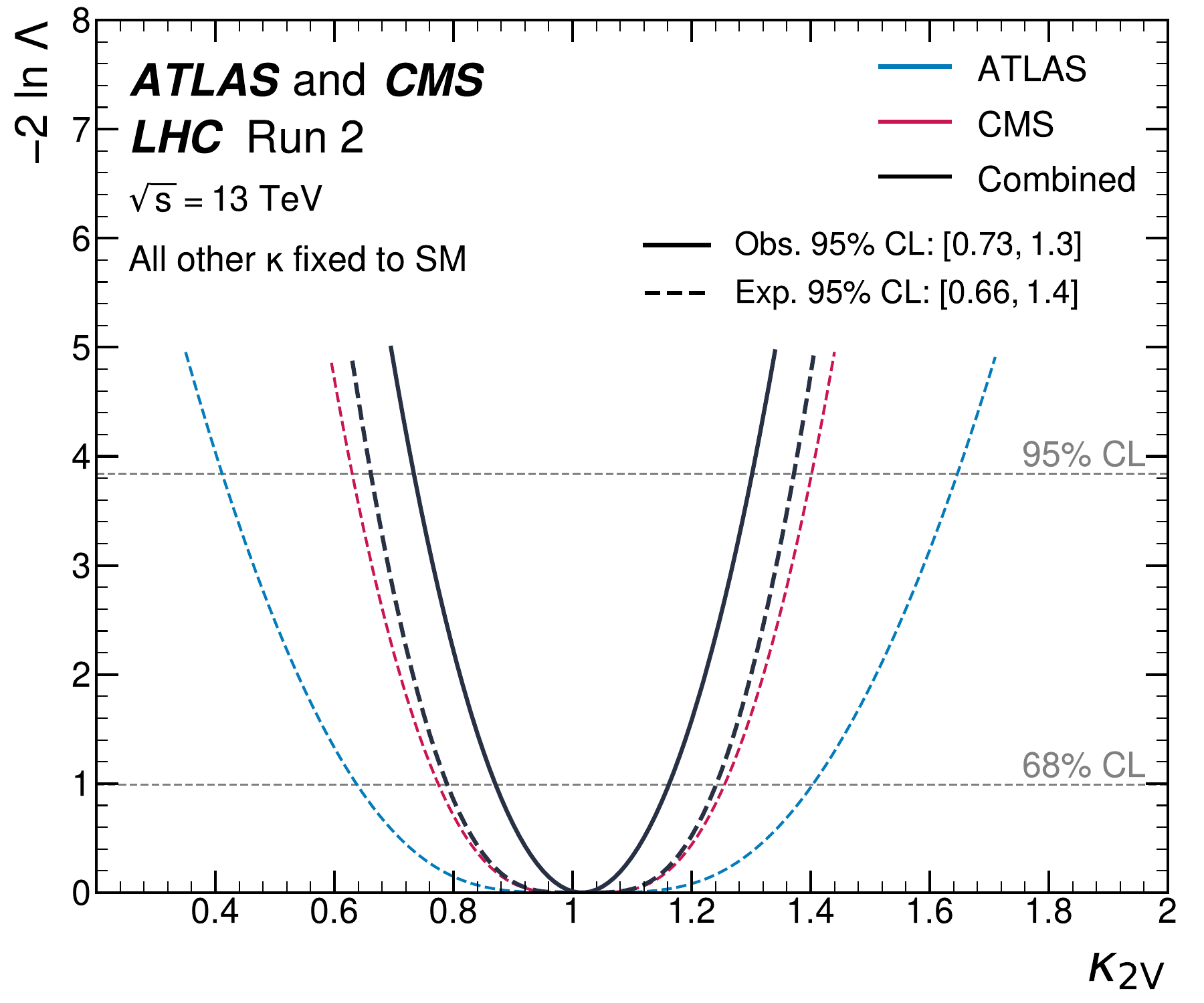}
\label{fig:likelihoodscan:kvv}
}
\caption{
Expected negative log-likelihood ($-2\ln\Lambda$) values as functions of (a) \klambda and (b) \kvv for ATLAS, CMS,  and the expected and observed values for the combination from both experiments.
All the other Higgs boson couplings are fixed to their SM predictions.
}
\label{fig:likelihoodscan}
\end{figure}

\begin{figure}[h]
\centering
\includegraphics[width=0.49\linewidth]{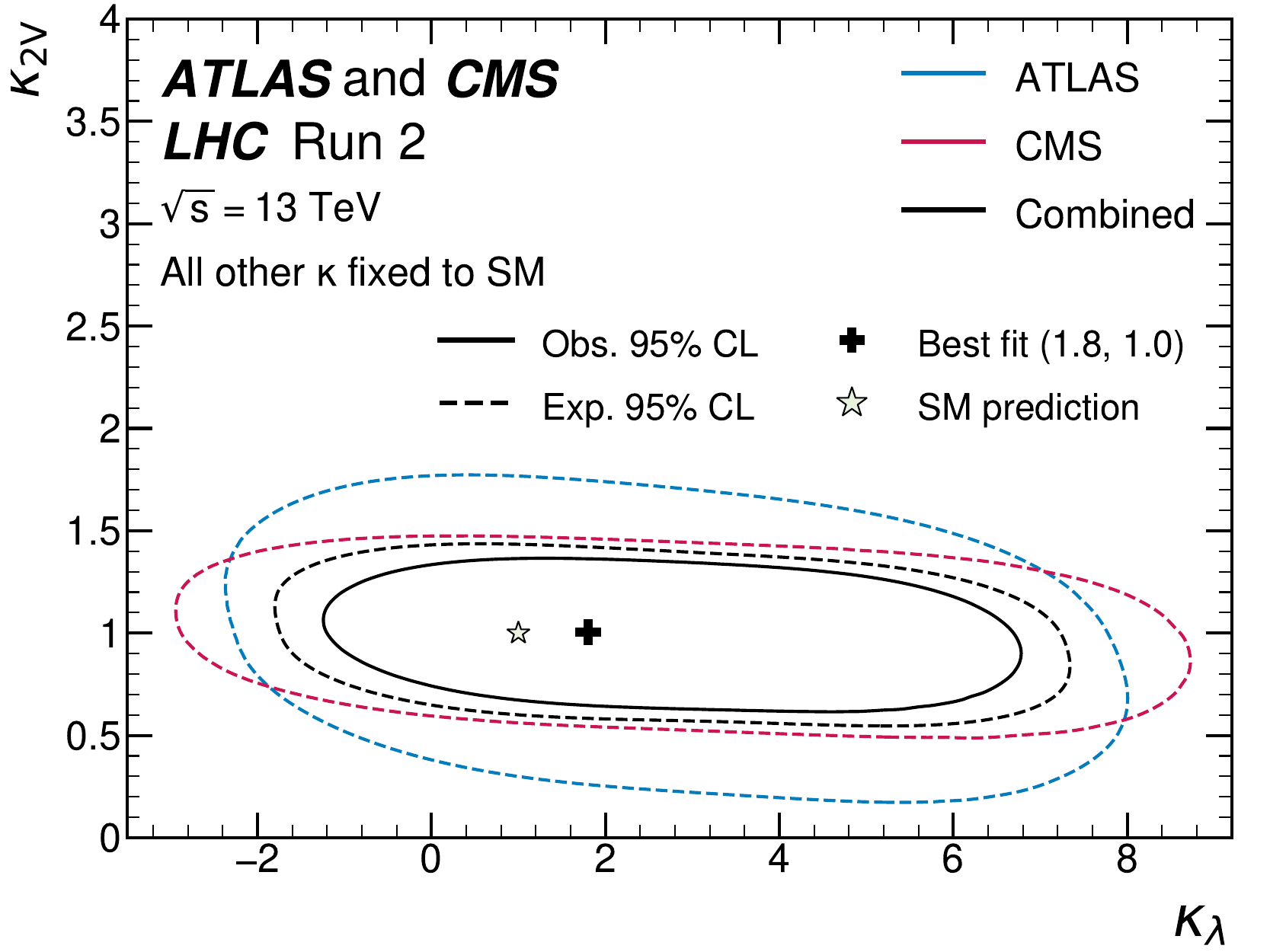}
\caption{Observed and expected 95\% CL contours for the simultaneous scan of the profile likelihood as a function of \klambda and \kvv. The expected constraints from the individual experiments are also shown. All the other Higgs boson couplings are fixed to their SM predictions.}
\label{fig:klkvvscan}
\end{figure}

Figure~\ref{fig:likelihoodscan} shows the expected and observed likelihood values as a function of \klambda and \kvv obtained from individual fits to the ATLAS and CMS data, and their combination.
Only the Higgs boson coupling considered in the corresponding fit is allowed to vary while the others are fixed to their SM predictions.
The observed combined constraints at 95\% CL on the coupling modifier values are
$-0.71 < \klambda < 6.1$ and
$0.73 < \kvv < 1.3$,
while the expected constraints are
$-1.3 < \klambda < 6.7$ and
$0.66 < \kvv < 1.4$ assuming the presence of the SM \HH signal.
The observed best-fit values for the coupling modifiers from the individual fits are
$\hat{\kappa}_\lambda = 1.8^{+2.8}_{-1.5}$
and
$\hat{\kappa}_\text{2V} = 1.02^{+0.14}_{-0.15}$
while the expected values are
$\hat{\kappa}_\lambda^\text{exp} = 1.0^{+4.2}_{-1.3}$
and
$\hat{\kappa}_\text{2V}^\text{exp} = 1.00^{+0.24}_{-0.21}$.
The channels with leading sensitivity on \klambda are \HHbbgg and \HHbbtt, while the \kvv result is driven by the \bbbb analyses in the Lorentz-boosted topologies because of the increase in the average \PTH in scenarios with $\kvv\neq 1$, resulting in much stronger sensitivities to these signals compared to the SM VBF \HH production.
The difference between the ATLAS and CMS CL intervals arises from the sensitivities of these channels in each experiment.
The smaller observed \kvv interval relative to the expected one is primarily driven by the ATLAS \bbbb analysis, which reports a deficit of events in the Lorentz-boosted topology, thereby driving the fit towards $\kvv$ values near one, where the VBF \HH production cross section is smallest.
The presented combination improves the expected sensitivity on \klambda by 10\% and on \kvv by 8\% with respect to the best results to date, as given by the ATLAS and CMS individual combinations, respectively.
The 95\% CL contours, obtained from a fit where \klambda and \kvv are allowed to vary simultaneously, are shown in Figure~\ref{fig:klkvvscan}.
The best-fit point in this plane corresponds to 1.8 for \klambda and 1.0 for \kvv.
All results are compatible with the SM predictions.

Table~\ref{tab:inputnalysis} presents a summary of the individual sensitivities of the channels considered, of the single experiment combinations, and of the ATLAS and CMS combination presented in this Letter.
Results from the individual channels and the single experiment combinations have been recomputed with respect to the original references following the modifications to the signal normalization and $\klambda$ dependence for the \HH and \PH processes described above.

\begin{table}[htbp]
\centering
\caption{Summary of the observed and expected results from the individual input analyses considered, the single experiment combinations, and the LHC combination presented in this Letter. The upper limit on $\mu_\HH=\sigma_{\HH}/\sigma_{\HH}^\text{SM}$ and the constraints on \klambda and \kvv are reported. All values quoted are defined at 95\% CL. The expected limits on $\mu_\HH$ are computed assuming the absence of \HH signal. The references indicate the original publication where the corresponding analyses are documented, and the results were recomputed following the modifications to input analyses described in the text.
}
\label{tab:inputnalysis}
\scalebox{0.93}{
\begin{tabular}{lllllllll}
\toprule
& \multirow{2}{*}{Channel} & \multirow{2}{*}{Details}
& \multicolumn{3}{c}{Expected} & \multicolumn{3}{c}{Observed} \\
\cmidrule(r{0.25em}){4-6} \cmidrule(l{0.25em}){7-9}
& &
& $\mu_{\HH}$ & $\klambda$ & $\kvv$
& $\mu_{\HH}$ & $\klambda$ & $\kvv$ \\
\midrule
\multirow{6}{*}{\rotatebox[origin=c]{90}{ATLAS}}
& \HHbbbb & \cite{HDBS-2019-29,HDBS-2022-02}
& $< 8.1$ & [$-5.3$, 12] & [$0.38$, 1.7]
& $< 5.4$ & [$-3.4$, 11] & [$0.55$, 1.5] \\
&  \HHbbtt & \cite{HDBS-2019-27}
& $< 3.3$ & [$-2.6$, 9.2] & [$-0.24$, 2.4]
& $< 5.8$ & [$-3.3$, 9.1] & [$-0.51$, 2.7] \\
&  \HHbbgg & \cite{HDBS-2021-10}
& $< 5.2$ & [$-3.0$, 7.9] & [$-1.1$, 3.3]
& $< 4.1$ & [$-1.6$, 7.0] & [$-0.48$, 2.7] \\
&  \HHbbll & \cite{HDBS-2019-02}
& $< 14$  & [$-11$, 17] & [$-0.50$, 2.7]
& $< 9.6$ & [$-6.5$, 13] & [$-0.19$, 2.4] \\
& Multilepton & \cite{HDBS-2019-04}
& $< 11$  & [$-4.5$, 9.6] & [$-1.9$, 4.1]
& $< 17$  & [$-6.4$, 12] & [$-2.5$, 4.7] \\
\cmidrule(lr){2-9}
& Combined & \cite{ATLAS:2024ish}
& $< 2.4$ & [$-1.7$, 7.2] & [0.41, 1.7]
& $< 2.9$ & [$-1.3$, 7.2] & [0.57, 1.5] \\
\midrule[\heavyrulewidth]
\multirow{6}{*}{\rotatebox[origin=c]{90}{CMS}}
&\HHbbbb &\cite{Tumasyan:2801882,arXiv:2205.06667}
& $< 4.3$ & [$-4.6$, 12] & [0.63, 1.4]
& $< 7.0$ & [$-5.0$, 12] & [0.66, 1.4] \\
&\HHbbtt & \cite{arXiv:2206.09401}
& $< 5.4$ & [$-4.2$, 11] & [$-0.64$, 2.8]
& $< 3.5$ & [$-1.9$, 8.9] & [$-0.32$, 2.5] \\
&  \HHbbgg & \cite{Sirunyan:2745738}
& $< 5.7$ & [$-3.5$, 8.8] & [$-0.94$, 3.1]
& $< 8.7$ & [$-3.5$, 8.0] & [$-1.4$, 3.6] \\
&\HHbbWW & \cite{CMS-HIG-21-005}
& $< 19$  & [$-9.4$, 16] & [$-1.4$, 3.5]
& $< 15$  & [$-6.1$, 13] & [$-1.0$, 3.1]  \\
&Multilepton &\cite{CMS:2022kdx}
& $< 20$  & [$-8.0$, 12] & [$-2.5$, 4.6]
& $< 22$  & [$-5.8$, 10] & [$-3.4$, 5.6] \\
\cmidrule(lr){2-9}
& Combined & \cite{CMS_combination}
& $< 2.6$ & [$-2.2$, 8.0] & [0.63, 1.4]
& $< 3.5$ & [$-1.4$, 6.6] & [0.66, 1.4] \\
\midrule
\multicolumn{3}{l}{ATLAS + CMS combined}
& $< 1.7$ & [$-1.3$, 6.7] & [0.66, 1.4]
& $< 2.5$ & [$-0.71$, 6.1] & [0.73, 1.3] \\
\bottomrule

\end{tabular}
}
\end{table}

In summary, this Letter presents the first combination of ATLAS and CMS searches for Higgs boson pair production.
The searches are performed using the LHC Run-2 data sets with integrated luminosities ranging between 126 and 140\fbinv.
The upper limit at 95\% CL on the total \HH signal strength, defined as the ratio of the measured cross section to the SM prediction, corresponds to 2.5, with an expected value of 1.7 (2.8) in the absence (presence) of SM \HH production.
The best-fit of the signal strength is observed to be
$0.8^{+0.9}_{-0.7}$, corresponding to an observed significance of 1.1 standard deviations, while the expected significance is 1.3 standard deviations.
The observed 95\% CL constraints on the \klambda and $\kvv$ coupling modifiers are
$-0.71 < \klambda < 6.1$ ($-1.3 < \klambda < 6.7$ expected) and
$0.73 < \kvv < 1.3$ ($0.66 < \kvv < 1.4$ expected).
The results of the combination are compatible within their uncertainties with the SM predictions and represent the most comprehensive and most sensitive constraints on \HH production to date.


\FloatBarrier

\section*{Acknowledgements}
%

%

%
%

%
%

We thank CERN for the very successful operation of the LHC and its injectors, as well as the support staff at
CERN and at our institutions worldwide without whom ATLAS could not be operated efficiently.

The crucial computing support from all WLCG partners is acknowledged gratefully, in particular from CERN, the ATLAS Tier-1 facilities at TRIUMF/SFU (Canada), NDGF (Denmark, Norway, Sweden), CC-IN2P3 (France), KIT/GridKA (Germany), INFN-CNAF (Italy), NL-T1 (Netherlands), PIC (Spain), RAL (UK) and BNL (USA), the Tier-2 facilities worldwide and large non-WLCG resource providers. Major contributors of computing resources are listed in Ref.~\cite{ATL-SOFT-PUB-2026-001}.

We gratefully acknowledge the support of ANPCyT, Argentina; YerPhI, Armenia; ARC, Australia; BMWFW and FWF, Austria; ANAS, Azerbaijan; CNPq and FAPESP, Brazil; NSERC, NRC and CFI, Canada; CERN; ANID, Chile; CAS, MOST and NSFC, China; Minciencias, Colombia; MEYS CR, Czech Republic; DNRF and DNSRC, Denmark; IN2P3-CNRS and CEA-DRF/IRFU, France; SRNSFG, Georgia; BMFTR, HGF and MPG, Germany; GSRI, Greece; RGC and Hong Kong SAR, China; ICHEP and Academy of Sciences and Humanities, Israel; INFN, Italy; MEXT and JSPS, Japan; CNRST, Morocco; NWO, Netherlands; RCN, Norway; MNiSW, Poland; FCT, Portugal; MNE/IFA, Romania; MSTDI, Serbia; MSSR, Slovakia; ARIS and MVZI, Slovenia; DSI/NRF, South Africa; MICIU/AEI, Spain; SRC and Wallenberg Foundation, Sweden; SERI, SNSF and Cantons of Bern and Geneva, Switzerland; NSTC, Taipei; TENMAK, T\"urkiye; STFC/UKRI, United Kingdom; DOE and NSF, United States of America.

Individual groups and members have received support from BCKDF, CANARIE, CRC and DRAC, Canada; CERN-CZ, FORTE and PRIMUS, Czech Republic; COST, ERC, ERDF, Horizon 2020 and Marie Sk{\l}odowska-Curie Actions, European Union; Investissements d'Avenir Labex, Investissements d'Avenir Idex and ANR, France; DFG and AvH Foundation, Germany; Herakleitos, Thales and Aristeia programmes co-financed by EU-ESF and the Greek NSRF, Greece; BSF-NSF and MINERVA, Israel; NCN and NAWA, Poland; La Caixa Banking Foundation, CERCA Programme Generalitat de Catalunya and PROMETEO and GenT Programmes Generalitat Valenciana, Spain; G\"{o}ran Gustafssons Stiftelse, Sweden; The Royal Society and Leverhulme Trust, United Kingdom; United States of America.

In addition, individual members wish to acknowledge support from Chile: Agencia Nacional de Investigaci\'on y Desarrollo (ANID FONDECYT reg. 1230987, FONDECYT 1230812, FONDECYT 1240864, Fondecyt 3240661, Fondecyt Regular 1240721); China: Chinese Ministry of Science and Technology (MOST-2023YFA1605700, MOST-2023YFA1609300), National Natural Science Foundation of China (NSFC - 12175119, NSFC 12275265); Czech Republic: Czech Science Foundation (GACR - 24-11373S), Ministry of Education Youth and Sports (ERC-CZ-LL2327, FORTE CZ.02.01.01/00/22\_008/0004632), PRIMUS Research Programme (PRIMUS/21/SCI/017); EU: H2020 European Research Council (ERC - 101002463); European Union: European Research Council (BARD No. 101116429, ERC - 948254, ERC 101089007), European Regional Development Fund (HE COFUND GA No.101081355, ERDF), European Union, Future Artificial Intelligence Research (FAIR-NextGenerationEU PE00000013), Marie Sklodowska-Curie Actions (GAP-101168829); France: Agence Nationale de la Recherche (ANR-21-CE31-0013, ANR-21-CE31-0022, ANR-22-EDIR-0002, ANR-24-CE31-0504-01); Germany: Deutsche Forschungsgemeinschaft (DFG - 469666862); China: Research Grants Council (GRF); Italy: Ministero dell'Università e della Ricerca (NextGenEU 153D23001490006 M4C2.1.1, NextGenEU I53D23000820006 M4C2.1.1, NextGenEU I53D23001490006 M4C2.1.1, SOE2024\_0000023); Japan: Japan Society for the Promotion of Science (JSPS KAKENHI JP22H01227, JSPS KAKENHI JP22H04944, JSPS KAKENHI JP22KK0227, JSPS KAKENHI JP24K23939, JSPS KAKENHI JP24KK0251, JSPS KAKENHI JP25H00650, JSPS KAKENHI JP25H01291, JSPS KAKENHI JP25K01023); Norway: Research Council of Norway (RCN-314472); Poland: Ministry of Science and Higher Education (IDUB AGH, POB8, D4 no 9722), Polish National Science Centre (NCN 2021/42/E/ST2/00350, NCN OPUS 2023/51/B/ST2/02507, NCN OPUS nr 2022/47/B/ST2/03059, NCN UMO-2019/34/E/ST2/00393, UMO-2022/47/O/ST2/00148, UMO-2023/49/B/ST2/04085, UMO-2023/51/B/ST2/00920, UMO-2024/53/N/ST2/00869); Spain: Agencia de Gestión de Ayudas Universitarias y de Investigación (AGAUR - 2023 BP 00141), Generalitat Valenciana (ASFAE/2022/008), Ministry of Science and Innovation (RYC2019-028510-I, RYC2020-030254-I, RYC2021-031273-I, RYC2022-038164-I), Ministerio de Ciencia, Innovación y Universidades/Agencia Estatal de Investigaci\'on (PID2022-142604OB-C22); Sweden: Carl Trygger Foundation (Carl Trygger Foundation CTS 22:2312), Swedish Research Council (Swedish Research Council 2023-04654, VR 2021-03651, VR 2022-03845, VR 2022-04683, VR 2023-03403, VR 2024-05451), Knut and Alice Wallenberg Foundation (KAW 2018.0458, KAW 2023.0366); Switzerland: Swiss National Science Foundation (SNSF - PCEFP2\_194658); United Kingdom: The Binks Trust, Royal Society (NIF-R1-231091); United States of America: U.S. Department of Energy (ECA DE-AC02-76SF00515), John Templeton Foundation (John Templeton Foundation 63206), Neubauer Family Foundation.

%
%


We congratulate our colleagues in the CERN accelerator departments for the excellent performance of the LHC and thank the technical and administrative staffs at CERN and at other CMS institutes for their contributions to the success of the CMS effort. In addition, we gratefully acknowledge the computing centers and personnel of the Worldwide LHC Computing Grid and other centers for delivering so effectively the computing infrastructure essential to our analyses. Finally, we acknowledge the enduring support for the construction and operation of the LHC, the CMS detector, and the supporting computing infrastructure provided by the following funding agencies: SC (Armenia), BMBWF and FWF (Austria); FNRS and FWO (Belgium); CNPq, CAPES, FAPERJ, FAPERGS, and FAPESP (Brazil); MES and BNSF (Bulgaria); CERN; CAS, MoST, and NSFC (China); MINCIENCIAS (Colombia); MSES and CSF (Croatia); RIF (Cyprus); SENESCYT (Ecuador); ERC PRG, TARISTU24-TK10 and MoER TK202 (Estonia); Academy of Finland, MEC, and HIP (Finland); CEA and CNRS/IN2P3 (France); SRNSF (Georgia); BMFTR, DFG, and HGF (Germany); GSRI (Greece); NKFIH (Hungary); DAE and DST (India); IPM (Iran); SFI (Ireland); INFN (Italy); MSIT and NRF (Republic of Korea); MES (Latvia); LMTLT (Lithuania); MOE and UM (Malaysia); BUAP, CINVESTAV, CONACYT, LNS, SEP, and UASLP-FAI (Mexico); MOS (Montenegro); MBIE (New Zealand); PAEC (Pakistan); MES, NSC, and NAWA (Poland); FCT (Portugal); MESTD (Serbia); MICIU/AEI and PCTI (Spain); MOSTR (Sri Lanka); Swiss Funding Agencies (Switzerland); MST (Taipei); MHESI (Thailand); TUBITAK and TENMAK (T\"{u}rkiye); NASU (Ukraine); STFC (United Kingdom); DOE and NSF (USA).


%

%
%
%
%
%
%

%
%
%

\printbibliography
\clearpage
 
\begin{flushleft}
\hypersetup{urlcolor=black}
{\Large The ATLAS Collaboration}

\bigskip

\AtlasOrcid[0000-0002-6665-4934]{G.~Aad}$^\textrm{\scriptsize 103}$,
\AtlasOrcid[0000-0001-7616-1554]{E.~Aakvaag}$^\textrm{\scriptsize 17}$,
\AtlasOrcid[0000-0002-5888-2734]{B.~Abbott}$^\textrm{\scriptsize 123}$,
\AtlasOrcid[0000-0002-0287-5869]{S.~Abdelhameed}$^\textrm{\scriptsize 119a}$,
\AtlasOrcid[0000-0002-1002-1652]{K.~Abeling}$^\textrm{\scriptsize 55}$,
\AtlasOrcid[0000-0001-5763-2760]{N.J.~Abicht}$^\textrm{\scriptsize 49}$,
\AtlasOrcid[0000-0002-8496-9294]{S.H.~Abidi}$^\textrm{\scriptsize 30}$,
\AtlasOrcid[0009-0003-6578-220X]{M.~Aboelela}$^\textrm{\scriptsize 45}$,
\AtlasOrcid[0000-0002-9987-2292]{A.~Aboulhorma}$^\textrm{\scriptsize 36e}$,
\AtlasOrcid[0000-0001-5329-6640]{H.~Abramowicz}$^\textrm{\scriptsize 156}$
% ...
\bigskip
\\

% ...
$^{17}$Department for Physics and Technology, University of Bergen, Bergen; Norway.\\
% ...
$^{30}$Physics Department, Brookhaven National Laboratory, Upton NY; United States of America.\\
% ...
$^{36}$$^{(a)}$Facult\'e des Sciences Ain Chock, Universit\'e Hassan II de Casablanca;$^{(b)}$Facult\'{e} des Sciences, Universit\'{e} Ibn-Tofail, K\'{e}nitra;$^{(c)}$Facult\'e des Sciences Semlalia, Universit\'e Cadi Ayyad, LPHEA-Marrakech;$^{(d)}$LPMR, Facult\'e des Sciences, Universit\'e Mohamed Premier, Oujda;$^{(e)}$Facult\'e des sciences, Universit\'e Mohammed V, Rabat;$^{(f)}$Institute of Applied Physics, Mohammed VI Polytechnic University, Ben Guerir; Morocco.\\
% ...
$^{45}$Physics Department, Southern Methodist University, Dallas TX; United States of America.\\
$^{46}$National Centre for Scientific Research "Demokritos", Agia Paraskevi; Greece.\\
% ...
$^{49}$Fakult\"{a}t Physik, Technische Universit{\"a}t Dortmund, Dortmund; Germany.\\
% ...
$^{55}$II. Physikalisches Institut, Georg-August-Universit\"{a}t G\"ottingen, G\"ottingen; Germany.\\
% ...
$^{103}$CPPM, Aix-Marseille Universit\'e, CNRS/IN2P3, Marseille; France.\\
% ...
$^{119}$$^{(a)}$New York University Abu Dhabi, Abu Dhabi;$^{(b)}$United Arab Emirates University, Al Ain; United Arab Emirates.\\
% ...
$^{123}$Homer L. Dodge Department of Physics and Astronomy, University of Oklahoma, Norman OK; United States of America.\\
% ...
$^{156}$Raymond and Beverly Sackler School of Physics and Astronomy, Tel Aviv University, Tel Aviv; Israel.\\
$^{157}$Department of Physics, Aristotle University of Thessaloniki, Thessaloniki; Greece.\\
% ...
$^{a}$ Also at Affiliated with an institute formerly covered by a cooperation agreement with CERN.\\
$^{b}$ Also at An-Najah National University, Nablus; Palestine.\\
$^{c}$ Also at Borough of Manhattan Community College, City University of New York, New York NY; United States of America.\\
$^{d}$ Also at Center for Interdisciplinary Research and Innovation (CIRI-AUTH), Thessaloniki; Greece.\\
$^{e}$ Also at Centre of Physics of the Universities of Minho and Porto (CF-UM-UP); Portugal.\\
$^{f}$ Also at CERN, Geneva; Switzerland.\\
% ...
$^{i}$ Also at Department of Financial and Management Engineering, University of the Aegean, Chios; Greece.\\
% ...
$^{*}$ Deceased

\end{flushleft}


%

\begin{flushleft}
\hypersetup{urlcolor=black}{\Large The CMS Collaboration}

\bigskip

A.~Hayrapetyan$^{1}$,
V.~Makarenko\orcidlink{0000-0002-8406-8605}
% ...
$^{1,b}$,
W.~Adam\orcidlink{0000-0001-9099-4341}
% ...
$^{2,c}$,
N.~Krammer\orcidlink{0000-0002-0548-0985}
% ...
$^{2,c}$,
R.~Sch\"{o}fbeck\orcidlink{0000-0002-2332-8784}$^{2,c}$,
M.~Shooshtari\orcidlink{0009-0004-8882-4887}
% ...
$^{2,c}$,
T.~Janssen\orcidlink{0000-0002-3998-4081}
% ...
$^{4,d}$,
G.P.~Van~Onsem\orcidlink{0000-0002-1664-2337}
% ...
$^{8,e}$,
A.~Vilela~Pereira\orcidlink{0000-0003-3177-4626}
% ...
$^{9,f}$,
M.~Costa~Reis\orcidlink{0000-0001-6892-7572}
% ...
$^{17,i}$,
H.S.~Chen\orcidlink{0000-0001-8672-8227}$^{17,i}$,
M.~Chen\orcidlink{0000-0003-0489-9669}$^{17,i}$,
Y.~Chen\orcidlink{0000-0002-4799-1636}$^{17}$,
Q.~Hou\orcidlink{0000-0002-1965-5918}$^{17}$,
X.~Hou$^{17}$,
F.~Iemmi\orcidlink{0000-0001-5911-4051}$^{17}$,
C.H.~Jiang$^{17}$,
H.~Liao\orcidlink{0000-0002-0124-6999}$^{17}$,
G.~Liu\orcidlink{0000-0001-7002-0937}$^{17}$,
Z.-A.~Liu\orcidlink{0000-0002-2896-1386}$^{17,j}$,
J.N.~Song$^{17,j}$,
S.~Song$^{17}$,
J.~Tao\orcidlink{0000-0003-2006-3490}$^{17}$,
C.~Wang$^{17,i}$,
J.~Wang\orcidlink{0000-0002-3103-1083}$^{17}$,
H.~Zhang\orcidlink{0000-0001-8843-5209}$^{17}$,
J.~Zhao\orcidlink{0000-0001-8365-7726}$^{17}$,
A.~Agapitos\orcidlink{0000-0002-8953-1232}
% ...
$^{30}$,
K.~Christoforou\orcidlink{0000-0003-2205-1100}$^{30}$,
S.~Konstantinou\orcidlink{0000-0003-0408-7636}$^{30}$,
C.~Leonidou\orcidlink{0009-0008-6993-2005}$^{30}$,
L.~Paizanos\orcidlink{0009-0007-7907-3526}$^{30}$,
F.~Ptochos\orcidlink{0000-0002-3432-3452}$^{30}$,
P.A.~Razis\orcidlink{0000-0002-4855-0162}$^{30}$,
H.~Rykaczewski$^{30}$,
H.~Saka\orcidlink{0000-0001-7616-2573}$^{30}$,
A.~Stepennov\orcidlink{0000-0001-7747-6582}$^{30}$,
M.~Finger\orcidlink{0000-0002-7828-9970}$^{31,a}$,
M.~Finger~Jr.\orcidlink{0000-0003-3155-2484}
% ...
$^{36}$,
A.~Milieva\orcidlink{0000-0001-5975-7305}$^{36}$,
K.~Osterberg\orcidlink{0000-0003-4807-0414}$^{36}$,
M.~Voutilainen\orcidlink{0000-0002-5200-6477}$^{36}$,
F.~Garcia\orcidlink{0000-0002-4023-7964}
% ...
$^{45}$,
S.~Consuegra~Rodr\'{i}guez\orcidlink{0000-0002-1383-1837}$^{45}$,
L.~Feld\orcidlink{0000-0001-9813-8646}$^{45}$,
K.~Klein\orcidlink{0000-0002-1546-7880}$^{45}$,
M.~Lipinski\orcidlink{0000-0002-6839-0063}$^{45}$,
P.~Nattland\orcidlink{0000-0001-6594-3569}$^{45}$,
V.~Oppenl\"{a}nder$^{45}$,
A.~Pauls\orcidlink{0000-0002-8117-5376}$^{45}$,
D.~P\'{e}rez~Ad\'{a}n\orcidlink{0000-0003-3416-0726}$^{45}$,
N.~R\"{o}wert\orcidlink{0000-0002-4745-5470}$^{45}$,
C.~Daumann$^{46}$,
S.~Diekmann\orcidlink{0009-0004-8867-0881}$^{46}$,
N.~Eich\orcidlink{0000-0001-9494-4317}$^{46}$,
D.~Eliseev\orcidlink{0000-0001-5844-8156}$^{46}$,
F.~Engelke\orcidlink{0000-0002-9288-8144}$^{46}$,
J.~Erdmann\orcidlink{0000-0002-8073-2740}$^{46}$,
M.~Erdmann\orcidlink{0000-0002-1653-1303}$^{46}$,
B.~Fischer\orcidlink{0000-0002-3900-3482}$^{46}$,
T.~Hebbeker\orcidlink{0000-0002-9736-266X}$^{46}$,
K.~Hoepfner\orcidlink{0000-0002-2008-8148}$^{46}$,
A.~Jung\orcidlink{0000-0002-2511-1490}$^{46}$,
N.~Kumar\orcidlink{0000-0001-5484-2447}$^{46}$,
M.y.~Lee\orcidlink{0000-0002-4430-1695}$^{46}$,
F.~Mausolf\orcidlink{0000-0003-2479-8419}$^{46}$,
M.~Merschmeyer\orcidlink{0000-0003-2081-7141}$^{46}$,
A.~Meyer\orcidlink{0000-0001-9598-6623}$^{46}$,
A.~Pozdnyakov\orcidlink{0000-0003-3478-9081}$^{46}$,
W.~Redjeb\orcidlink{0000-0001-9794-8292}$^{46}$,
H.~Reithler\orcidlink{0000-0003-4409-702X}$^{46}$,
U.~Sarkar\orcidlink{0000-0002-9892-4601}$^{46}$,
V.~Sarkisovi\orcidlink{0000-0001-9430-5419}$^{46}$,
A.~Schmidt\orcidlink{0000-0003-2711-8984}$^{46}$,
C.~Seth$^{46}$,
A.~Sharma\orcidlink{0000-0002-5295-1460}$^{46}$,
J.L.~Spah\orcidlink{0000-0002-5215-3258}$^{46}$,
V.~Vaulin$^{46}$,
S.~Zaleski$^{46}$,
M.R.~Beckers\orcidlink{0000-0003-3611-474X}
% ...
$^{49}$,
M.~Antonello\orcidlink{0000-0001-9094-482X}$^{49}$,
S.~Bollweg$^{49}$,
M.~Bonanomi\orcidlink{0000-0003-3629-6264}$^{49}$,
L.~Ebeling$^{49}$,
K.~El~Morabit\orcidlink{0000-0001-5886-220X}$^{49}$,
Y.~Fischer\orcidlink{0000-0002-3184-1457}$^{49}$,
M.~Frahm$^{49}$,
E.~Garutti\orcidlink{0000-0003-0634-5539}$^{49}$,
A.~Grohsjean\orcidlink{0000-0003-0748-8494}$^{49}$,
A.A.~Guvenli\orcidlink{0000-0001-5251-9056}$^{49}$,
J.~Haller\orcidlink{0000-0001-9347-7657}$^{49}$,
D.~Hundhausen$^{49}$,
G.~Kasieczka\orcidlink{0000-0003-3457-2755}$^{49}$,
P.~Keicher\orcidlink{0000-0002-2001-2426}$^{49}$,
R.~Klanner\orcidlink{0000-0002-7004-9227}$^{49}$,
W.~Korcari\orcidlink{0000-0001-8017-5502}$^{49}$,
T.~Kramer\orcidlink{0000-0002-7004-0214}$^{49}$,
C.c.~Kuo$^{49}$,
F.~Labe\orcidlink{0000-0002-1870-9443}$^{49}$,
J.~Lange\orcidlink{0000-0001-7513-6330}$^{49}$,
A.~Lobanov\orcidlink{0000-0002-5376-0877}$^{49}$,
J.~Matthiesen$^{49}$,
L.~Moureaux\orcidlink{0000-0002-2310-9266}$^{49}$,
K.~Nikolopoulos\orcidlink{0000-0002-3048-489X}$^{49}$,
A.~Paasch\orcidlink{0000-0002-2208-5178}$^{49}$,
K.J.~Pena~Rodriguez\orcidlink{0000-0002-2877-9744}$^{49}$,
N.~Prouvost$^{49}$,
B.~Raciti\orcidlink{0009-0005-5995-6685}$^{49}$,
M.~Rieger\orcidlink{0000-0003-0797-2606}$^{49}$,
D.~Savoiu\orcidlink{0000-0001-6794-7475}$^{49}$,
P.~Schleper\orcidlink{0000-0001-5628-6827}$^{49}$,
M.~Schr\"{o}der\orcidlink{0000-0001-8058-9828}$^{49}$,
J.~Schwandt\orcidlink{0000-0002-0052-597X}$^{49}$,
M.~Sommerhalder\orcidlink{0000-0001-5746-7371}$^{49}$,
H.~Stadie\orcidlink{0000-0002-0513-8119}$^{49}$,
G.~Steinbr\"{u}ck\orcidlink{0000-0002-8355-2761}$^{49}$,
R.~Ward\orcidlink{0000-0001-5530-9919}$^{49}$,
B.~Wiederspan$^{49}$,
M.~Wolf\orcidlink{0000-0003-3002-2430}$^{49}$,
C.~Yede\orcidlink{0009-0002-3570-8132}$^{49}$,
A.~Brusamolino\orcidlink{0000-0002-5384-3357}
% ...
$^{55}$,
D.~Horvath\orcidlink{0000-0003-0091-477X}$^{55,bb,cc}$,
\'{A}.~Kadlecsik\orcidlink{0000-0001-5559-0106}$^{55}$,
C.~Lee\orcidlink{0000-0001-6113-0982}$^{55}$,
K.~M\'{a}rton$^{55}$,
A.J.~R\'{a}dl\orcidlink{0000-0001-8810-0388}$^{55,dd}$,
F.~Sikler\orcidlink{0000-0001-9608-3901}$^{55}$,
V.~Veszpremi\orcidlink{0000-0001-9783-0315}$^{55}$,
M.~Csan\'{a}d\orcidlink{0000-0002-3154-6925}
% ...
$^{87a,a}$,
C.~Carrivale$^{87a,87b}$,
D.~Ciangottini\orcidlink{0000-0002-0843-4108}
% ...
$^{90a,90c,a}$,
N.~Bartosik\orcidlink{0000-0002-7196-2237}
% ...
$^{103}$,
Y.~Gharbia\orcidlink{0000-0002-0156-9448}$^{103}$,
F.~Alazemi\orcidlink{0009-0005-9257-3125}
% ...
$^{119}$,
K.~Pozniak\orcidlink{0000-0001-5426-1423}$^{119}$,
W.~Zabolotny\orcidlink{0000-0002-6833-4846}$^{119}$,
M.~Araujo\orcidlink{0000-0002-8152-3756}
% ...
$^{123}$,
J.~Alcaraz~Maestre\orcidlink{0000-0003-0914-7474}$^{123}$,
Cristina~F.~Bedoya\orcidlink{0000-0001-8057-9152}$^{123}$,
J.A.~Brochero~Cifuentes\orcidlink{0000-0003-2093-7856}$^{123}$,
Oliver~M.~Carretero\orcidlink{0000-0002-6342-6215}$^{123}$,
M.~Cepeda\orcidlink{0000-0002-6076-4083}$^{123}$,
M.~Cerrada\orcidlink{0000-0003-0112-1691}$^{123}$,
N.~Colino\orcidlink{0000-0002-3656-0259}$^{123}$,
B.~De~La~Cruz\orcidlink{0000-0001-9057-5614}$^{123}$,
A.~Delgado~Peris\orcidlink{0000-0002-8511-7958}$^{123}$,
A.~Escalante~Del~Valle\orcidlink{0000-0002-9702-6359}$^{123}$,
D.~Fern\'{a}ndez~Del~Val\orcidlink{0000-0003-2346-1590}$^{123}$,
J.P.~Fern\'{a}ndez~Ramos\orcidlink{0000-0002-0122-313X}$^{123}$,
J.~Flix\orcidlink{0000-0003-2688-8047}$^{123}$,
M.C.~Fouz\orcidlink{0000-0003-2950-976X}$^{123}$,
M.~Gonzalez~Hernandez\orcidlink{0009-0007-2290-1909}$^{123}$,
O.~Gonzalez~Lopez\orcidlink{0000-0002-4532-6464}$^{123}$,
S.~Goy~Lopez\orcidlink{0000-0001-6508-5090}$^{123}$,
J.M.~Hernandez\orcidlink{0000-0001-6436-7547}$^{123}$,
M.I.~Josa\orcidlink{0000-0002-4985-6964}$^{123}$,
J.~Llorente~Merino\orcidlink{0000-0003-0027-7969}$^{123}$,
C.~Martin~Perez\orcidlink{0000-0003-1581-6152}$^{123}$,
E.~Martin~Viscasillas\orcidlink{0000-0001-8808-4533}$^{123}$,
D.~Moran\orcidlink{0000-0002-1941-9333}$^{123}$,
C.~M.~Morcillo~Perez\orcidlink{0000-0001-9634-848X}$^{123}$,
\'{A}.~Navarro~Tobar\orcidlink{0000-0003-3606-1780}$^{123}$,
R.~Paz~Herrera\orcidlink{0000-0002-5875-0969}$^{123}$,
A.~P\'{e}rez-Calero~Yzquierdo\orcidlink{0000-0003-3036-7965}$^{123}$,
J.~Puerta~Pelayo\orcidlink{0000-0001-7390-1457}$^{123}$,
I.~Redondo\orcidlink{0000-0003-3737-4121}$^{123}$,
J.~Vazquez~Escobar\orcidlink{0000-0002-7533-2283}$^{123}$,
J.F.~de~Troc\'{o}niz\orcidlink{0000-0002-0798-9806}
% ...
$^{156}$,
S.~Baradia\orcidlink{0000-0001-9860-7262}$^{156}$,
B.~Barton\orcidlink{0000-0003-4390-5881}$^{156}$,
R.~Breedon\orcidlink{0000-0001-5314-7581}$^{156}$,
H.~Cai\orcidlink{0000-0002-5759-0297}$^{156}$,
M.~Calderon~De~La~Barca~Sanchez\orcidlink{0000-0001-9835-4349}$^{156}$,
E.~Cannaert$^{156}$,
M.~Chertok\orcidlink{0000-0002-2729-6273}$^{156}$,
M.~Citron\orcidlink{0000-0001-6250-8465}$^{156}$,
J.~Conway\orcidlink{0000-0003-2719-5779}$^{156}$,
P.T.~Cox\orcidlink{0000-0003-1218-2828}$^{156}$,
F.~Eble\orcidlink{0009-0002-0638-3447}$^{156}$,
R.~Erbacher\orcidlink{0000-0001-7170-8944}$^{156}$,
O.~Kukral\orcidlink{0009-0007-3858-6659}$^{156}$,
G.~Mocellin\orcidlink{0000-0002-1531-3478}$^{156}$,
S.~Ostrom\orcidlink{0000-0002-5895-5155}$^{156}$,
I.~Salazar~Segovia$^{156}$,
J.S.~Tafoya~Vargas\orcidlink{0000-0002-0703-4452}$^{156}$,
W.~Wei\orcidlink{0000-0003-4221-1802}$^{156}$,
S.~Yoo\orcidlink{0000-0001-5912-548X}$^{156}$,
K.~Adamidis$^{157}$,
M.~Bachtis\orcidlink{0000-0003-3110-0701}$^{157}$,
D.~Campos$^{157}$,
R.~Cousins\orcidlink{0000-0002-5963-0467}$^{157}$,
S.~Crossley\orcidlink{0009-0008-8410-8807}$^{157}$,
G.~Flores~Avila\orcidlink{0000-0001-8375-6492}$^{157}$,
J.~Hauser\orcidlink{0000-0002-9781-4873}$^{157}$,
M.~Ignatenko\orcidlink{0000-0001-8258-5863}$^{157}$,
M.A.~Iqbal\orcidlink{0000-0001-8664-1949}$^{157}$,
T.~Lam\orcidlink{0000-0002-0862-7348}$^{157}$,
Y.f.~Lo\orcidlink{0000-0001-5213-0518}$^{157}$,
E.~Manca\orcidlink{0000-0001-8946-655X}$^{157}$,
A.~Nunez~Del~Prado\orcidlink{0000-0001-7927-3287}$^{157}$,
D.~Saltzberg\orcidlink{0000-0003-0658-9146}$^{157}$,
V.~Valuev\orcidlink{0000-0002-0783-6703}$^{157}$,
R.~Clare\orcidlink{0000-0003-3293-5305}
% ...
$^{188,a}$,
R.~Demina\orcidlink{0000-0002-7852-167X}
% ...
$^{17}$Institute of High Energy Physics, Beijing, China\\
% ...
$^{30}$University of Cyprus, Nicosia, Cyprus\\
% ...
$^{36}$Department of Physics, University of Helsinki, Helsinki, Finland\\
% ...
$^{45}$RWTH Aachen University, I. Physikalisches Institut, Aachen, Germany\\
$^{46}$RWTH Aachen University, III. Physikalisches Institut A, Aachen, Germany\\
% ...
$^{49}$University of Hamburg, Hamburg, Germany\\
% ...
$^{55}$HUN-REN Wigner Research Centre for Physics, Budapest, Hungary\\
% ...
$^{103}$College of Engineering and Technology, American University of the Middle East (AUM), Dasman, Kuwait\\
% ...
$^{119}$Warsaw University of Technology, Warsaw, Poland\\
% ...
$^{123}$Centro de Investigaciones Energ\'{e}ticas Medioambientales y Tecnol\'{o}gicas (CIEMAT), Madrid, Spain\\
% ...
$^{156}$University of California, Davis, Davis, California, USA\\
$^{157}$University of California, Los Angeles, California, USA\\
% ...
$^{a}$Deceased\\
$^{b}$Also at Yerevan State University, Yerevan, Armenia\\
$^{c}$Also at TU Wien, Vienna, Austria\\
$^{d}$Also at Ghent University, Ghent, Belgium\\
$^{e}$Also at FACAMP - Faculdades de Campinas, Sao Paulo, Brazil\\
$^{f}$Also at Universidade Estadual de Campinas, Campinas, Brazil\\
% ...
$^{i}$Also at University of Chinese Academy of Sciences, Beijing, China\\
% ...
\end{flushleft}


%

%
%

\clearpage

\end{document}